\documentclass[prd,aps,10pt,twocolumn,nofootinbib,superscriptaddress,showpacs]{revtex4-1}

\usepackage{amssymb,amsmath}
\usepackage{xcolor}
\usepackage[utf8]{inputenc}

\usepackage{graphicx}

\usepackage[makeroom]{cancel}
\usepackage[caption=false]{subfig}
\usepackage[colorlinks=true,
            citecolor=green,
            linkcolor=red,
            filecolor=cyan,
            urlcolor=magenta,
            backref=false]{hyperref}

\usepackage{enumitem}

\newcommand{\dd}{\mbox{d}}

\begin{document}

\title{Regular black hole from a confined spin connection in Poincar\'e gauge gravity}

% \title{Black hole singularity resolution with a confined spin connection}
% \title{Asymptotically free cosmological constant in Poincar\'e gauge gravity gives rise to finite renormalization group-improved Schwarzschild--de\,Sitter black hole}

% \title{Renormalization group improvement in Poincar\'e gauge gravity:\\generating a regular Schwarzschild--de\,Sitter black hole}

\author{Jens Boos}
\email{jboos@wm.edu}
\affiliation{High Energy Theory Group, Department of Physics, William \& Mary, Williamsburg, VA 23187-8795, United States}

\date{\today}

\begin{abstract}
Within the asymptotic safety program, it is possible to construct renormalization group (RG) improved spacetimes by replacing the gravitational coupling $G$ by its running counterpart $G(k)$, and subsequently identifying the RG scale $k$ with a physical distance scale. This procedure has been used to construct a regular Schwarzschild geometry, but it fails in the presence of a cosmological constant. This can only be avoided if the dimensionless cosmological constant has a trivial ultraviolet fixed point, but so far no such scenario has been encountered in quantum general relativity (with or without matter). In this Letter we provide a possible solution to this problem. In Poincar\'e gauge gravity an effective cosmological constant arises naturally, and if the non-Abelian Lorentz spin connection is asymptotically free, it generates a trivial ultraviolet fixed point for this cosmological constant. We thereby tentatively propose a nonsingular black hole consistent with the principles of asymptotic safety, embedded in Poincar\'e gauge gravity.
\end{abstract}

\maketitle

\section{Introduction}

General relativity's prediction of black hole singularities is perhaps one of the strongest indications that this theory is not ultraviolet complete. While it is generally believed that quantum gravity effects will resolve spacetime singularities \cite{Bojowald:2007ky},\footnote{For somewhat contrasting views, see Refs.~\cite{Horowitz:1995ta,Bousso:2022tdb}.} providing such a mechanism in all generality has proven difficult. Instead, considerable attention has been devoted to so-called \emph{nonsigular black hole spacetimes} \cite{Bardeen:1968,Dymnikova:1992ux,Hayward:2005gi}; see Ref.~\cite{Frolov:2016pav} for a recent review. Typically, these  geometries are derived from a singular seed metric (say, the Schwarzschild black hole) that is supplemented by a new physics parameter $\ell$ that regulates the curvature singularity; in the limiting case of $\ell \rightarrow 0$ one again recovers the initial seed metric. It should be noted that the new physics modification is \textit{ad hoc}, and such regular metrics do not solve the vacuum Einstein equations; moreover, it has been argued that some regular black hole geometries cannot arise from a local, well-defined least action principle \cite{Giacchini:2021pmr,Knorr:2022kqp}.

Since Weinberg's proposal of asymptotic safety \cite{Weinberg:1979}, the advent of functional renormalization group methods \cite{Polchinski:1983gv,Wetterich:1992yh,Morris:1993qb,Reuter:1996cp,Berges:2000ew} eventually gave rise to the field of asymptotically safe quantum gravity \cite{Niedermaier:2006wt,Percacci:2011fr}; see Refs.~\cite{Donoghue:2019clr,Bonanno:2020bil} for critical reflections. Within this field, it has since been pointed out that replacing the Newtonian coupling $G$ with its renormalization group (RG) induced running counterpart $G(k)$ can be used to generate regular black hole models, provided the momentum scale $k$ is suitably identified with a spatial distance scale \cite{Bonanno:2000ep,Koch:2013owa,Moti:2018uvl,Adeifeoba:2018ydh,Bosma:2019aiu,Bonanno:2020fgp,Bonanno:2023rzk}, which is sometimes referred to as ``cutoff identification.''\footnote{The fact that asymptotically free general relativity may resolve black hole singularities had been pointed out earlier in Ref.~\cite{Frolov:1988vj}.} Perhaps somewhat surprisingly, in the context of quantum electrodynamics, this procedure gives the correct 1-loop expression for the Uehling potential from the classical Coulomb potential, but fails to give accurate results at next-to-leading order \cite{Bonanno:2000ep}. Hence, in the gravitational setting, this proocedure may capture the essential leading order quantum effects in gravity. It should be noted, however, that this approach is certainly not unique since the renormalization scale $k$ needs to be identified with a physical distance scale, which in most cases is inherently coordinate dependent and therefore not unique; for a suggested iterative procedure that does not violate general covariance, see Platania \cite{Platania:2019kyx}.

More generally, RG improvement has been known to yield regular geometries in the context of \emph{asymptotically flat} black hole spacetimes. When extended to asymptotically (anti-)de\,Sitter spacetimes, the introduction of a cosmological constant and its RG running spoils the regularity of the spacetime and reintroduces curvature singularities. This can only be avoided if the dimensionless cosmological constant has a trivial UV fixed point \cite{Koch:2013owa}. Such an asymptotically free cosmological constant, however, is not expected in the functional renormalization group analysis of Einstein--Hilbert truncations (with our without matter) \cite{Adeifeoba:2018ydh}, and, while it can be stipulated as an additional assumption, may also merely point to a fundamental incompleteness of Einstein gravity.

In this Letter we would like to investigate how these conclusions are affected in an alternative description of gravity, so-called Poincar\'e gauge gravity \cite{Hehl:1976kj,Blagojevic:2013}; for a recent introduction see Obukhov \cite{Obukhov:2022khx}. This class of gravitational theories is consistent with the Newtonian limit, features many exact black hole solutions, and slightly widens the geometric arena to not only accomodate spacetime curvature, but also spacetime torsion. It has been argued that such gravity theories include new massive short-range gauge bosons \cite{Boos:2016cey}, affecting the UV behavior of gravity and potentially regularizing matter sources \cite{Poplawski:2009su}. A complete study of the particle spectrum of Poincar\'e gauge gravity has been developed, identifying viable sectors devoid of tachyons and ghosts \cite{Karananas:2014pxa,Blagojevic:2018dpz,Lin:2018awc,Lin:2019ugq,Barker:2022kdk}; it appears as if the inclusion of torsion partially resolves the well-known ghost problem of quadratic Einstein gravity \cite{Stelle:1976gc,Stelle:1977ry}. For renormalizability properties of Poincar\'e gauge gravity at one loop, see Melichev and Percacci \cite{Melichev:2023lwj}. Since the RG flow in Poincar\'e gauge gravity differs from that of Einstein gravity, in this Letter we study the Schwarzschild--de\,Sitter solution of Poincar\'e gauge gravity, and determine, in an analogous study to \cite{Adeifeoba:2018ydh}, what kind of constraints are imposed on the couplings' fixed points by demanding a nonsingular geometry. We find, rather surprisingly, that within Poincar\'e gauge gravity a regular Schwarzschild--de\,Sitter geometry appears to follow from a confined (or ``asymptotically free'') spin connection. While detailed RG analyses are required to determine the precise running of the couplings in Poincar\'e gauge gravity, it can be argued group-theoretically that at one loop the non-Abelian gravity sector is indeed asymptotically free, which implies a regular Schwarzschild--de\,Sitter geometry.

This Letter is organized as follows: In Sec.~\ref{sec:pgg} we briefly present the main features of Poincar\'e gauge gravity and its Schwarzschild--de\,Sitter solution. In Sec.~\ref{sec:rg} we model the RG running of the gravitational couplings and extract regularity conditions. In Sec.~\ref{sec:af} we argue that the additional gravitational coupling exhibits asymptotic freedom, thereby generating a nonsingular, asymptotically (anti-)de\,Sitter black hole spacetime. We summarize and conclude in Sec.~\ref{sec:conclusion}.

\section{Exact black hole solution in Poincar\'e gauge gravity}
\label{sec:pgg}

The origins of Poincar\'e gauge gravity date back to the work of Cartan, Utiyama, Sciama,  Kibble, and others \cite{Blagojevic:2013}. Taking the conservation of energy momentum and total angular momentum in relativistic field theories (defined in flat Minkowski spacetime, in the absence of gravity) as initial ingredients, gauging the global Poincar\'e transformations then gives rise to a model of gravity (in the same sense that, say, gauging the translations gives rise to supergravity \cite{VanNieuwenhuizen:1981ae}). Since the Poincar\'e group $P_4$ has a semidirect product structure of translations $T_4$ and Lorentz rotations $SO(1,3)$, $P_4 = T_4 \rtimes SO(1,3)$, one is required to introduce ten gauge potentials: four gauge potentials for the gauged translations ($e^\mu_i$, the ``vielbein''), and six gauge potentials for the gauged Lorentz rotations ($\Gamma_{i\alpha}{}^\beta$, the ``spin connection''). However, the semidirect product structure leads to ``crosstalk'' between the translational and the rotational gauge sectors, which makes Poincar\'e gauge gravity rather different from Yang--Mills theories of product gauge groups.

\subsection{Poincar\'e gauge theory in a nutshell}
The spacetime metric, in the framework of Poincar\'e gauge gravity, is a derived quantity and is given by
\begin{align}
g{}_{\mu\nu} = e_\mu^i e_\nu^j g{}_{ij} = \text{diag}(-1,1,1,1) \, .
\end{align}
Here, Greek (Latin) indices refer to pseudo-orthogonal (coordinate) frames; the object $e_i^\mu$ is called the vielbein and serves as the translational gauge potential; it is sometimes also called \emph{soldering form} since it can be used to convert frame tensors into coordinate tensors, and vice versa. Hence, $g{}_{ij}$ denotes the usual metric of the manifold, and $g{}_{\mu\nu}$ encodes the diagonal, Minkowski-like metric in the pseudo-orthogonal basis. The spin connection, serving as the rotational gauge potential, is no such tensor; its coordinate expression $\Gamma{}_{ij}{}^k$ and its frame expression $\Gamma_{i\alpha}{}^\beta$ are related via
\begin{align}
\Gamma{}_{ij}{}^k = e_j^\alpha e^k_\beta \Gamma{}_{i\alpha}{}^\beta + e^k_\gamma \partial_i e_j^\gamma \, .
\end{align}
The two field strengths of Poincar\'e gauge gravity are torsion and curvature, respectively:
\begin{align}
\begin{split}
T_{ij}{}^\alpha &= \partial_i e_j^\alpha - \partial_j e_i^\alpha + \Gamma{}_{i\beta}{}^\alpha e_j^\beta - \Gamma{}_{j\beta}{}^\alpha e_i^\beta  \, , \\
R{}_{ij\alpha}{}^\beta &= \partial_i \Gamma{}_{j\alpha}{}^\beta - \partial_j \Gamma{}_{i\alpha}{}^\beta + \Gamma_{i\gamma}{}^\beta \Gamma_{j\alpha}{}^\gamma - \Gamma_{j\gamma}{}^\beta \Gamma_{i\alpha}{}^\gamma \, .
\end{split}
\end{align}
Since $T{}_{ij}{}^\alpha$ and $R{}_{ij}{}^\alpha{}_\beta$ are Poincar\'e tensors, one can equivalently work in the coordinate basis,
\begin{align}
T{}_{ij}{}^k \equiv e{}^k_\alpha T{}_{ij}{}^\alpha \, , \quad R{}_{ijk}{}^l \equiv e{}^\alpha_k e{}^l_\beta R{}_{ij\alpha}{}^\beta \, .
\end{align}
Therein, they may be interpreted as the Poincar\'e-valued holonomy for vector fields \cite{Anandan:1994tx}:
\begin{align}
\left[\nabla_i, \nabla_j\right]V^k = R{}_{ija}{}^k V^a - T{}_{ij}{}^a \nabla_a V^k \, ,
\end{align}
that is, curvature mediates a Lorentz rotation around an infinitesimal loop, and torsion mediates a translation. Spacetimes with non-vanishing torsion and curvature are called Riemann--Cartan spacetimes, and they reduce to Riemannian spacetimes in the limit of vanishing torsion.

To compare and contrast Riemann--Cartan spacetimes from purely Riemannian spacetimes it can be helpful to split off purely non-Riemannian contributions to the geometry. To that end, one defines the contortion tensor $K_{ij}{}^k$ that encapsulates the non-Riemannian part of the spin connection:
\begin{align}
\begin{split}
K_{ij}{}^k &\equiv \Gamma{}_{ij}{}^k - \widetilde{\Gamma}_{ij}{}^k \\
&= -\frac12 \left( T{}_{ij}{}^k + T^{k}{}_{ij} + T{}^k{}_{ji} \right)
\end{split}
\end{align}
Not surprisingly, the contortion tensor is proportional to the torsion tensor. In other words, the Levi-Civita connection encountered in general relativity is simply $\widetilde{\Gamma}_{ij}{}^k$, that is, it is the torsion-independent part of the connection (and in what follows, all purely Riemannian objects shall be decorated with a tilde). We can use this quantity to define the purely Riemannian curvature tensor,
\begin{align}
\widetilde{R}{}_{ijk}{}^l &\equiv \partial_i \widetilde{\Gamma}{}_{jk}{}^l - \partial_j \widetilde{\Gamma}{}_{ik}{}^l + \widetilde{\Gamma}{}_{im}{}^l \widetilde{\Gamma}{}_{jk}{}^m - \widetilde{\Gamma}{}_{jm}{}^l \widetilde{\Gamma}{}_{ik}{}^m \, .
\end{align}
Last, derived from this object, one may consider the Ricci tensor $\widetilde{R}_{ij} \equiv \widetilde{R}{}_{iaj}{}^a$, the Ricci scalar $\widetilde{R} \equiv\widetilde{R}^i{}_i$, and the Weyl tensor
\begin{align}
\begin{split}
\widetilde{C}{}_{ijkl} \equiv \widetilde{R}{}_{ijkl} &- g{}_{i[k} \widetilde{R}{}_{l]j} + g{}_{j[k} \widetilde{R}{}_{l]i} + \frac13 \widetilde{R} g{}_{i[k} g{}_{l]j} \, .
\end{split}
\end{align}
The presence of these additional curvature tensors shows the geometric richness of post-Riemannian geometries, and implies that the regularity conditions in such a setting may differ from those encountered within the Riemannian setting of general relativity.

The action of Poincar\'e gauge gravity $S_\text{PG}$ is viewed as a functional of the vielbein $e^\mu_i$ and the spin connection $\Gamma_{i\alpha}{}^\beta$, who are treated as independent variables, marking an important distinction to the so-called Palatini formalism (which constitutes a formal rewriting of general relativity in terms of frame variables). Denoting the matter action as $S_m$, the two field equations of Poincar\'e gauge gravity are obtained via the variations
\begin{align}
\begin{split}
\frac{\delta S_\text{PG}[e^\mu_i, \Gamma_{i\alpha}{}^\beta]}{\delta e_j^\nu} &= \mathfrak{T}^j{}_\nu \, , \\
\frac{\delta S_\text{PG}[e^\mu_i, \Gamma_{i\alpha}{}^\beta]}{\delta \Gamma_{k\rho}{}^\sigma} &= \mathfrak{S}^{k\rho}{}_\sigma \, ,
\end{split}
\end{align}
where $\mathfrak{T}^j{}_\nu \equiv \delta S_\text{m}/\delta e_j^\nu$ is the energy-momentum tensor, and $\mathfrak{S}^{k\rho}{}_\sigma \equiv \delta S_\text{m}/\delta \Gamma_{k\rho}{}^\sigma$ is the spin-angular momentum tensor (that is, the sum of orbital angular momentum as well as intrinsic spin). Correspondingly, a solution in Poincar\'e gauge gravity is always specified by the tuple $(e^\mu_i, \Gamma_{i\alpha}{}^\beta)$, or, equivalently, by the tuple $(e^\mu_i, T_{ij}{}^\alpha)$; it is also possible to specify the tuple $(g{}_{ij}, T_{ij}{}^\alpha)$. The last tuple suggests that once a solution to the field equations has been found, one may think of it as a spacetime metric, accompanied by torsion contributions.

\subsection{Schwarzschild--de\,Sitter solution}
So far, we only described the general structure of Poincar\'e gauge gravity, which is an entire class of gravitational theories. In order to fix the theory under consideration, let us consider the following gravitational Lagrangian:
\begin{align}
\label{eq:vdh-lagrangian}
\mathcal{L} = \frac{1}{16\pi} \left[ \frac{1}{G} \left( T{}_{ij}{}^k T{}^{ij}{}_k - T{}_{ji}{}^j T{}^{ki}{}_k \right) - \frac{1}{\rho} R{}_{ij}{}^{kl} R{}^{ij}{}_{kl} \right] \, .
\end{align}
This model has been originally proposed by von der Heyde \cite{VonDerHeyde:1976}, and has been met with considerable attention in the literature; see Obukhov \cite{Obukhov:2019fti} and references therein. It closely resembles the shape of a traditional Yang--Mills-type Lagrangian, and does not contain the linear Ricci curvature scalar, nor a cosmological constant.

The field equations are a bit lengthy, and they are displayed in \cite{Obukhov:2019fti}. In the vacuum case, $\mathfrak{T}^j{}_\nu = \mathfrak{S}^{k\rho}{}_\sigma = 0$, they are solved by the Baekler--Lee geometry \cite{Baekler:1981lkh,Lee:1983af} (see also Obukhov \cite{Obukhov:2019fti,Obukhov:2022khx}), which is also called the Schwarzschild--de\,Sitter solution of Poincar\'e gauge gravity:
\begin{align}
\begin{split}
\dd s^2 &= g{}_{ij}\dd x^i \dd x^j = -f \dd t^2 + \frac{\dd r^2}{f} + r^2\dd\Omega^2 \, , \\
f &= 1 - \frac{2Gm}{r} - \frac{\rho r^2}{4G} \, , \\
T{}_{\hat{1}\hat{0}}{}^{\hat{0}} &= T{}_{\hat{1}\hat{0}}{}^{\hat{1}} = T{}_{\hat{0}\hat{2}}{}^{\hat{2}} = T{}_{\hat{2}\hat{1}}{}^{\hat{2}} = T{}_{\hat{0}\hat{3}}{}^{\hat{3}} = T{}_{\hat{3}\hat{1}}{}^{\hat{3}} \equiv T \, , \\
T &= \frac{Gm}{f r^2} = \frac{Gm}{r^2 - 2Gmr - \frac{\rho \, r^4}{4G}} \, .
\end{split}
\end{align}
Here, the hatted indices $\hat{0}, \hat{1}, \dots$ label the frame components. For completeness, the vielbein is given by
\begin{align}
\begin{split}
(e^i_\mu) &= \text{diag}\left( f^{1/2}, f^{-1/2}, r, r\sin\theta \right) \, , \\
(e_i^\mu) &= \text{diag}\left( f^{-1/2}, f^{1/2}, \frac{1}{r}, \frac{1}{r\sin\theta} \right) \, .
\end{split}
\end{align}
Comparing the function $f(r)$ to the one giving rise to the Schwarzschild--de\,Sitter geometry \cite{Stephani:2003,Boos:2014hua} one can read off the effective cosmological constant of this model,
\begin{align}
\Lambda_\text{eff} = \frac{3\rho}{4G} \, .
\end{align}
The existence of such an effective cosmological constant presents a stark difference to general relativity, and such effective cosmological terms are not uncommon within Poincar\'e gauge gravity \cite{Hehl:1978yt}. Before moving on, let us understand the regularity properties of this geometry in a bit more detail.

To begin with, the metric function $f$ and the torsion function $T$ are singular at $r=0$. However, computing proper curvature and torsion invariants one finds the simple expressions
\begin{align}
\begin{split}
\label{eq:constant-curvature}
& R = \frac{3\rho}{G} \, , \\
& R_{ij} R^{ij} = \frac{9\rho^2}{4G^2} \, , \\
& C_{ijkl} = 0 \, , \\
& T{}_{ij}{}^k T{}^{ij}{}_k = 0 \, .
\end{split}
\end{align}
That is, viewed from a Riemann--Cartan perspective, the Schwarzschild--de\,Sitter solution features constant Ricci curvature; its Weyl tensor vanishes identically (which is a stark departure from general relativity), and its torsion tensor is null since it squares to zero, whereas its components are non-vanishing.

Taking instead the Riemannian curvature invariants derived solely from the Levi--Civita connection, one finds
\begin{align}
\begin{split}
& \widetilde{R} = \frac{3\rho}{G} \, , \\
& \widetilde{R}_{ij} \widetilde{R}^{ij} = \frac{9\rho^2}{4G^2} \, , \\
& \widetilde{C}_{ijkl} \widetilde{C}^{ijkl} = \frac{48G^2m^2}{r^6} \, .
\end{split}
\end{align}
While the expressions in the Ricci sector coincide with the Riemann--Cartan invariants, the purely Riemannian Weyl tensor invariant does not; in fact, it coincides with the expression one finds in general relativity for the Schwarzschild--de\,Sitter black hole. As $r \rightarrow 0$, this expression diverges. A necessary requirement for a regular geometry is that \emph{all} of its curvature invariants are bounded, and hence the Schwarzschild--de\,Sitter geometry of Poincar\'e gauge gravity is singular.

In what follows, let us address how this singularity can be lifted with concepts from asymptotic safety, applied to the running couplings $G$ and $\rho$ of the von der Heyde model \eqref{eq:vdh-lagrangian} in Poincar\'e gauge gravity.

\section{RG running and singularity resolution conditions}
\label{sec:rg}

Defining the RG scale dependent couplings $\rho(k)$ and $g(k) \equiv G(k)k^2$, we parametrize their running close to the fixed point as \cite{Adeifeoba:2018ydh}
\begin{align}
\begin{split}
\label{eq:scaling}
g(k) &= g_\star + g_1 \left(\frac{k}{M_\text{Pl}}\right)^{-\theta_1} + g_2 \left(\frac{k}{M_\text{Pl}}\right)^{-\theta_2} \\
&\equiv g_\star + \delta_g \, , \\
\rho(k) &= \rho_\star + \rho_1 \left(\frac{k}{M_\text{Pl}}\right)^{-\theta_1} + \rho_2 \left(\frac{k}{M_\text{Pl}}\right)^{-\theta_2} \\
&\equiv \rho_\star + \delta_\rho \, ,
\end{split}
\end{align}
where $\theta_1$ and $\theta_2$ are the critical exponents. Inserting the scale-dependent couplings into the metric function $f(r)$ one obtains ``neither fish nor fowl,''
\begin{align}
\label{eq:fff}
f_k(r) = 1 - \frac{2g(k)m}{k^2 r} + \frac{\rho(k) r^2 k^2}{2 g(k)} \, ,
\end{align}
that is, a quantity that is defined both in real space and momentum space simultaneously. Following the conventional approach, we now parametrize the cutoff identification via
\begin{align}
\label{eq:cutoff-identification}
k(r) \simeq \frac{\xi}{r} \left( \frac{G_0 m}{r} \right)^{\gamma-1} \, ,
\end{align}
where $\xi > 0$ is a free numerical parameter, $G_0 \equiv G(0)$, and $\gamma$ is a free parameter encountered in the literature; the choice $\gamma = 3/2$ is geometrically motivated \cite{Bonanno:2000ep}.

\subsection{Leading order}
Omitting (for now) the next-to-leading order scalings, inserting merely the fixed point values supplemented by \eqref{eq:cutoff-identification} and choosing $\gamma=3/2$ one finds
\begin{align}
f(r) \simeq 1 - \frac{G_0 m\xi^2 \rho_\star}{4 g_\star r} - \frac{2 g_\star r^2}{g_0 \xi^2} \, .
\end{align}
The quadratic Weyl invariant then takes the form
\begin{align}
\widetilde{C}^2 \equiv \widetilde{C}_{ijkl} \widetilde{C}^{ijkl} = \frac{3 G_0^2 m^2 \xi^4\rho_\star^2}{4 g_\star^2} \frac{1}{r^6} \, . 
\end{align}
Hence, a necessary requirement for the regularity of this spacetime is $\rho_\star = 0$. In contrast to previous work \cite{Adeifeoba:2018ydh}, where a necessary condition was a trivial ultraviolet fixed point for the dimensionless cosmological constant, here this condition is recast into the asytmptotic freedom of a dimensionless, non-Abelian coupling.

\subsection{Conditions on critical exponents}
Let us now work out constraints on the critical exponents $\theta_1$ and $\theta_2$ by analyzing the invariants $R$ and $\widetilde{C}^2$. For a general function $f(r)$ they take the form
\begin{align}
\begin{split}
R &= 1 - \frac{2f^2}{r^2} - \frac{8f'f}{r} - 2 (f')^2 - 2 f'' f \, , \\
\widetilde{C}^2 &= \frac{3}{4r^4} \left[ 1 - f^2 + 2 f' f - (f')^2r^2 - f''f r^2 \right]^2 \, ,
\end{split}
\end{align}
and parametrizing $f(r) = 1 + c \, r^\delta$ one has
\begin{align}
\begin{split}
%R &= -2c^2(1+3\delta+2\delta^2)r^{2\delta-2} - 2c(2 + 3\delta + \delta^2)r^{\delta-2} \, , \\
R &= -2c(\delta+1)\left[ c(2\delta+1)r^{2\delta-2} + (\delta + 2)r^{\delta-2} \right] \, , \\
\widetilde{C}^2 &= \frac{4c^2}{3}(\delta-1)^2\left[ (2\delta-1)r^{2\delta-2} + (\delta-2)r^{\delta-2} \right]^2 \, .
\end{split}
\end{align}
Observe that no choice of $\delta$ can simultaneously render both invariants zero or finite. As a consistency check, note that for the scaling $\delta = -1$ one recovers $R = 0$ (Schwarzschild behavior).\footnote{For $\delta=2$, however, one finds $R = -6c(4 + 5cr^2)$, which is not constant. This is not surprising, since in Poincar\'e gauge gravity only the curvature of the \emph{full} Schwarzschild--de\,Sitter geometry is constant, Eq.~\eqref{eq:constant-curvature}, which may be interpreted via the nontrivial interplay of curvature and torsion due to the semidirect product structure of the Poincar\'e group.} It follows that a necessary condition for a nonsingular geometry is
\begin{align}
\delta \ge 2 \, ,
\end{align}
in agreement with \cite{Adeifeoba:2018ydh}. Note, however, that as an exponent for a Riemann--Cartan geometry with torsion, there is no prior reason to assume agreement with general relativity. Inserting now the scaling relations \eqref{eq:scaling} into the renormalization group-improved function $f_k(r)$, Eq.~\eqref{eq:fff}, one has to leading order
\begin{align}
f_k(r) &= 1 - \frac{2m}{k^2r} \left( g_\star + \delta_g \right) - \frac{k^2r^2}{4}\frac{\rho_\star + \delta_\rho}{g_\star + \delta_g} \\
&\approx 1 - \frac{2m}{k^2r} \left( g_\star + \delta_g \right) - \frac{k^2r^2}{4}\left( \frac{\rho_\star}{g_\star} + \frac{\delta_\rho}{g_\star} - \frac{\rho_\star \delta_g}{g_\star^2} \right) \, . \nonumber
\end{align}
After substituting the cutoff identification \eqref{eq:cutoff-identification}, one finds
\begin{align}
\begin{split}
f(r) = 1 &+ c_1 \hat{r}^{2\gamma-1} + c_2 \hat{r}^{2-2\gamma} \\
&+ c_3 \hat{r}^{(\theta_1+2)\gamma-1} + c_4 \hat{r}^{(\theta_2+2)\gamma-1} \\
&+ c_5 \hat{r}^{(\theta_1-2)\gamma+2} + c_6 \hat{r}^{(\theta_2-2)\gamma+2} \, , \\[5pt]
\hat{r} &\equiv r/(G_0 m) \, .
\end{split}
\end{align}
The constants $c_1,\dots,c_6$ are given by
\begin{align}
\begin{split}
c_1 &= -\frac{2g_\star}{\xi^2} \left(\frac{m}{M_\text{Pl}}\right)^2 \, , \\
c_2 &= -\frac{\rho_\star \xi^2}{4 g_\star} \, , \\
c_3 &= -\frac{2 g_1}{\xi^{2+\theta_1}} \left(\frac{m}{M_\text{Pl}}\right)^{2+\theta_1} \, , \\
c_4 &= -\frac{2 g_2}{\xi^{2+\theta_2}} \left(\frac{m}{M_\text{Pl}}\right)^{2+\theta_2} \, , \\
c_5 &= -\frac{g_\star\rho_1 - g_1 \rho_\star }{4 g_\star^2 \xi^{\theta_1-2}} \left(\frac{m}{M_\text{Pl}}\right)^{\theta_1} \, , \\
c_6 &= -\frac{g_\star\rho_2 - g_2 \rho_\star }{4 g_\star^2 \xi^{\theta_2-2}} \left(\frac{m}{M_\text{Pl}}\right)^{\theta_2} \, ,
\end{split}
\end{align}
where we used the relation $G_0 \equiv 1/M_\text{Pl}^2$. We can now match the regularity condition $\delta \ge 2$ with the exponents of all terms. The $c_1$-term implies $\gamma \ge 3/2$, whereas the problematic $c_2$-term vanishes since $\rho_\star = 0$. The $c_3$ and $c_4$-terms imply that $\theta_{1,2} \ge 3/\gamma-2$, which, for finite $\gamma$, is always superseded by the conditions derived from the $c_5$ and $c_6$-terms: $\theta_{1,2} \ge 2$. Hence we find
\begin{align}
\gamma \ge \frac32 \, , \quad \theta_1 \ge 2 \, , \quad \theta_2 \ge 2 \, , \quad \rho_\star = 0 \, .
\end{align}
This extends previous results, obtained for general relativity \cite{Adeifeoba:2018ydh}, to Poincar\'e gauge gravity. Remarkably, the exponent thresholds $\gamma$, $\theta_1$, and $\theta_2$ agree between the two different gravity models. This is somewhat unexpected and may hint towards universality across different geometrical theories of gravity.

\section{Asymptotic freedom}
\label{sec:af}

Let us now argue that the dimensionless strong gravity coupling $\rho$ may indeed exhibit asymptotic freedom in Poincar\'e gauge gravity. Recalling its appearance in the gravitational Lagrangian \eqref{eq:vdh-lagrangian},
\begin{align}
\mathcal{L} \supset -\frac{1}{16\pi \rho} (R_{ijkl})^2 \, ,
\end{align}
we emphasize the similarity with Yang--Mills theory,
\begin{align}
\mathcal{L}_\text{YM} = -\frac{1}{4 g^2} \left( F^A_{ij} \right)^2 \, .
\end{align}
Indeed, $R_{ij\mu}{}^\nu$ and $F{}_{ij}^A$ are both curvature tensors. The former is the Lorentz curvature that takes values in the Lie algebra of $SO(1,3)$, and the latter is the well-known QCD field strength tensor taking values in the Lie algebra of $SU(N)$. It is well known that in a gauge theory with coupling $g$ the 1-loop pure gauge $\beta$-function takes the form \cite{Gross:1973id,Politzer:1973fx,Caswell:1974gg}
\begin{align}
\beta(g) = - \frac{g^3}{(4\pi)^2} \frac{11 \, C_2}{3} \, ,
\end{align}
where $C_2$ denotes the quadratic Casimir of the gauge group. That is, without matter, the asymptotic freedom at 1-loop is determined by the sign of $C_2$. Denoting the generators of local Lorentz transformations as $L_{\mu\nu}$, we follow Donoghue \cite{Donoghue:2016vck} and define
\begin{align}
\begin{split}
[L_{\mu\nu}, L_{\rho\sigma}] &= -2i( g_{\rho[\mu} L{}_{\nu]\sigma} - g_{\sigma[\mu} L{}_{\nu]\rho} ) \\
&\equiv f_{[\mu\nu][\rho\sigma][\alpha\beta]} L^{\alpha\beta} \, ,
\end{split}
\end{align}
where we read off
\begin{align}
f{}_{[\mu\nu][\rho\sigma][\alpha\beta]} = g{}_{\nu[\rho} g{}_{\sigma][\alpha} g{}_{\beta]\mu} - g{}_{\mu[\rho} g{}_{\sigma][\alpha} g{}_{\beta]\nu} \, .
\end{align}
Inserting this into
\begin{align}
\begin{split}
f{}_{[\mu\nu][\alpha\beta][\gamma\delta]} f{}^{[\rho\sigma][\alpha\beta][\gamma\delta]} &\equiv C_2 \delta{}^{[\rho\sigma]}_{[\mu\nu]} \\
&\equiv \frac{C_2}{2} \left( \delta{}^\rho_\mu \delta{}^\sigma_\nu - \delta{}^\sigma_\mu \delta{}^\rho_\nu \right) \, ,
\end{split}
\end{align}
one can read off $C_2 = 2$. Note that this result is independent of the metric signature, and also holds for $SO(4)$ in the purely Euclidean case since this defining relation only contains Kronecker deltas on the right-hand side.\footnote{This may appear somewhat surprising, since the Killing metric of a non-compact Lie group like $SO(1,3)$ is known to not be positive definite; however, at the level of the group Casimir this difference does not appear.}

Defining now $\rho' = \sqrt{4\pi\rho}$ we find
\begin{align}
\beta(\rho) = \frac{\rho'}{2\pi} \beta(\rho') = -\frac{11}{3\pi} \rho^2 \, ,
\end{align}
implying that in pure gravity the coupling $\rho$ indeed exhibits asymptotic freedom, $\rho_\star = 0$.

One may ask: does a similar argument hold for the coupling $G$ in Poincar\'e gauge gravity? The answer is negative, since the translational sector is Abelian,
\begin{align}
[\partial{}_i, \partial_j] = C{}_{ij}^k \partial_k \equiv 0 \, ,
\end{align}
and hence $C_2 = 0$ for that group.

Therefore, the non-Abelian (but \emph{not} the non-compact) nature of the $SO(1,3)$ spin connection is responsible for the ultraviolet fixed point, which is why Donoghue \cite{Donoghue:2016vck} refers to this behavior as ``confined,'' in analogy to the gluons in QCD. This way, the independent degrees of freedom present in an affine connection only play a role at high energies, while they are absent from large-distance, low-energy scenarios which has been argued from different perspective elsewhere \cite{Boos:2016cey}. For similar considerations in quadratic gravity, see Holdom and Ren \cite{Holdom:2015kbf}.

\section{Conclusions}
\label{sec:conclusion}

Renormalization group-improved black hole spacetimes are constructed by replacing all couplings by their running counterparts, and by then inserting a cutoff identification that relates the renormalization scale to a physical distance scale in the system under consideration. Typically, this then modifies the original black hole geometry close to the origin $r=0$, sometimes alleviating curvature singularities. If this procedure is applied to Einstein gravity and its Schwarzschild--de\,Sitter metric, demanding a regular metric implies the condition
\begin{align}
\lambda_\star = 0 \, ,
\end{align}
where $\lambda \equiv \Lambda/k^2$. Hence, the dimensionless cosmological constant must have a trivial ultraviolet fixed point \cite{Koch:2013owa}. This feature, however, has not yet emerged naturally in the functional renormalization group literature on quantum Einstein gravity (with our without matter) \cite{Adeifeoba:2018ydh}.

In this Letter we first showed that in Poincar\'e gauge gravity one encounters an effective cosmological constant $\Lambda_\text{eff} = 3\rho/(4G)$, where $G$ is Newton's coupling, and $\rho$ is the dimensionless coupling related to a curvature-squared piece of the Lagrangian. We then applied the renormalization group-improvement procedure to the Schwarzschild--de\,Sitter solution of Poincar\'e gauge gravity (which is accompanied by a non-trivial torsion field), wherein the requirement of $\lambda_\star = 0$ is replaced by the requirement
\begin{align}
\rho_\star = 0 \, .
\end{align}
We then argued that this condition is plausibly satisfied in Poincar\'e gauge gravity, using the general expression for pure-gauge $\beta$-functions at one loop to show that its sign is negative for $SO(1,3)$, implying an asymptotically free non-Abelian Lorentz sector in Poincar\'e gauge gravity. Such an interpretation is also attractive from a phenomenological point of view, since it would imply that the additional degrees of freedom encountered in this modified gravity theory are confined, and only relevant at high energies and short distances.

In this Letter we have hence constructed a consistent renormalization group-improved nonsingular black hole geometry in Poincar\'e gauge gravity, featuring an effective cosmological constant with a plausible trivial fixed point. This gives a motivation to extend the study of the full functional renormalization group from higher-curvature and higher-derivative Einstein gravity to extended, non-Riemannian geometries, and we hope to address this in more detail in the future. First steps appear to have been completed by Pagani and Percacci \cite{Pagani:2015ema}; however, their torsion and nonmetricity fields have been assumed non-propagating, and hence their considerations do not apply to the approach followed in the present Letter. At any rate, it appears that geometrical departures from purely Riemannian spacetimes can potentially solve some of the problems encountered in quantum Einstein gravity, without spoiling desirable infrared behaviors of gravity.\\

\section*{Acknowledgments}
I am grateful for support by the National Science Foundation under grants PHY-1819575 and PHY-2112460.

% \vfill \hfill \textit{rgbh-v6.tex, jb, Aug 22, 2023.}

% \appendix

\end{document}